\documentclass[aps,prl,twocolumn,superscriptaddress,nofootinbib]{revtex4-2}

\usepackage{amsmath,amssymb}
\usepackage{graphicx}
\usepackage{xcolor}

\begin{document}

\title{Lorentz-Covariant Spectral Bounds from Thermal Quantum Field Theory}

\author{Alisher Sanetullaev}
\email{a.sanetullaev@newuu.uz}
\affiliation{New Uzbekistan University}

\author{Sarbinaz Bazarbaeva}
\affiliation{New Uzbekistan University}

\author{Marhabo Beymamatova}
\affiliation{New Uzbekistan University}

\author{Shokir Tursunov}
\affiliation{New Uzbekistan University}

\date{\today}

\begin{abstract}
We derive rigorous Lorentz-covariant bounds on relaxation spectra directly from the analytic structure of retarded Green's functions in thermal quantum field theory, using only causality, unitarity, the Kubo--Martin--Schwinger condition, and Lorentz covariance --- without reference to any specific dynamical model. A single rest-frame quasinormal pole is generically smeared into a continuum of excitations in boosted frames, with width set by the maximal signal velocity. We prove that the non-hydrodynamic gap $\Gamma_\text{gap}$ transforms as $\tilde{\Gamma}_\text{gap} \geq \Gamma_\text{gap}/[\gamma(1 + v\, v_\text{max})]$, and that the convergence radius of the hydrodynamic gradient expansion satisfies $\tilde{k}_c \in [k_c/\gamma(1+v\,v_s),\, k_c/\gamma(1-v\,v_s)]$ under a boost of velocity $v$. We verify the bounds by a numerical quasinormal-mode computation in the $\mathcal{N}=4$ super-Yang-Mills plasma: the leading boosted pole moves \emph{deeper} into the complex plane --- the observed relaxation rate increases with boost velocity, in sharp contrast to naive time dilation --- while respecting the bound throughout. The results apply non-perturbatively to the quark-gluon plasma, neutron star merger dynamics, and quantum critical systems.
\end{abstract}

\maketitle

\emph{Introduction.}---The real-time collective dynamics of quantum many-body systems at finite temperature is governed by the pole structure of retarded Green's functions in the complex frequency plane. Hydrodynamic poles near the origin encode long-lived collective modes; non-hydrodynamic poles encode transient microscopic relaxation \cite{KovtunStarinets2005,Grozdanov2019complex}. How this pole structure transforms between inertial frames is central to relativistic hydrodynamics \cite{Romatschke2019,HoultKovtun2020}, heavy-ion phenomenology \cite{HeinzSnellings2013,Gale2013}, and the interpretation of holographic transport \cite{Policastro2001,KSS2005}.

Recently, Gavassino \cite{Gavassino2026PRL,Gavassino2026arXiv} proved that in linearized relativistic (kinetic or rheological) theories with an Onsager-type symmetry, Lorentz boosts generically split a single rest-frame relaxation mode into a continuum of excitations, with width set by the maximal signal speed, and derived rigorous bounds confining the boosted relaxation spectrum in terms of rest-frame spectral data alone. Those proofs operate at the level of classical linearized equations with Hermitian structure. Here we establish analogous bounds in full thermal quantum field theory (QFT), replacing the model-specific symmetry assumptions with the universal properties of retarded two-point functions: causality, unitarity, the Kubo--Martin--Schwinger (KMS) condition, and Lorentz covariance. Our bounds are therefore non-perturbative and hold for strongly coupled systems --- including holographic quasinormal spectra, which lie outside the symmetry class of Refs.~\cite{Gavassino2026PRL,Gavassino2026arXiv} and whose covariant treatment was posed there as an open problem. Full proofs and extended analysis appear in the companion paper \cite{companion}.

The central results are: (i)~a spectral smearing theorem showing how rest-frame poles broaden under boosts; (ii)~a covariant lower bound on the non-hydrodynamic gap; (iii)~a covariant upper bound on the maximal relaxation rate; and (iv)~covariant bounds on the convergence radius of the hydrodynamic gradient expansion. We verify these in holographic examples and discuss applications to the quark-gluon plasma (QGP) and neutron star mergers.

\emph{Setup: retarded Green's functions and quasinormal modes.}---Let $\mathcal{O}(x)$ be a local Hermitian bosonic operator in a relativistic QFT at temperature $T = 1/\beta$ and chemical potential $\mu$. The retarded Green's function
\begin{equation*}
G^R(\omega,\mathbf{k}) = \int d^4x\, e^{i\omega t - i\mathbf{k}\cdot\mathbf{x}}(-i)\theta(t)\langle[\mathcal{O}(x),\mathcal{O}(0)]\rangle_\beta
\end{equation*}
is analytic for $\operatorname{Im}(\omega)>0$ by causality, and admits the Kramers--Kronig representation
\begin{equation*}
G^R(\omega,\mathbf{k}) = \int_{-\infty}^\infty d\omega'\,\frac{\rho(\omega',\mathbf{k})}{\omega'-\omega-i0^+},
\end{equation*}
where $\rho(\omega,\mathbf{k}) = -\frac{1}{\pi}\operatorname{Im} G^R(\omega+i0^+,\mathbf{k}) \geq 0$ for $\omega > 0$ by unitarity. The KMS condition fixes the Wightman function via $G^>(\omega,\mathbf{k}) = 2\pi\rho(\omega,\mathbf{k})/(1-e^{-\beta\omega})$ (for an operator carrying $U(1)$ charge $q$, $\omega \to \omega - q\mu$).

The quasinormal mode (QNM) frequencies $\{\omega_n(\mathbf{k})\}$ are poles of $G^R$ in the lower half $\omega$-plane. Hydrodynamic modes satisfy $\omega_n(\mathbf{k})\to 0$ as $\mathbf{k}\to 0$; non-hydrodynamic modes have a gap
\begin{equation*}
\Gamma_\text{gap} \equiv \min_{n\in\text{non-hydro}} \bigl(-\operatorname{Im}[\omega_n(\mathbf{0})]\bigr) > 0.
\end{equation*}
The convergence radius $k_c$ of the hydrodynamic gradient expansion $\omega_\text{hydro}(\mathbf{k}) = \sum_n c_n |\mathbf{k}|^n$ is set by the collision of a hydrodynamic pole with the nearest non-hydrodynamic pole in the complex $k$-plane \cite{Grozdanov2019PRL,Heller2023}.

\emph{Spectral bounds under Lorentz boosts.}---For a Lorentz boost with velocity $\mathbf{v}$ [$\gamma = (1-v^2)^{-1/2}$], four-momentum transforms as $\tilde{\omega} = \gamma(\omega - \mathbf{v}\cdot\mathbf{k})$ and $\tilde{\mathbf{k}} = \mathbf{k} + (\gamma-1)\hat{v}(\hat{v}\cdot\mathbf{k}) - \gamma\mathbf{v}\omega$. For a scalar operator, Lorentz covariance gives $\tilde{G}^R(\tilde{p}) = G^R(\Lambda^{-1}\tilde{p})$, and thus
\begin{equation}
\tilde{\rho}(\tilde{\omega},\tilde{\mathbf{k}}) = \rho\!\left(\frac{\tilde{\omega}-v\tilde{k}_\parallel}{\gamma},\, \tilde{k}_\perp,\, \frac{\tilde{k}_\parallel - v\tilde{\omega}}{\gamma}\right),
\label{eq:rhotransform}
\end{equation}
for a longitudinal boost with $\tilde{k}_\parallel = \hat{v}\cdot\tilde{\mathbf{k}}$. Equation~\eqref{eq:rhotransform} is exact; it shows that the boosted spectral weight at frequency $\tilde{\omega}$ is sourced by the rest-frame spectral function at boosted spatial momentum $\tilde{k}_\parallel^\text{rest} = (\tilde{k}_\parallel - v\tilde{\omega})/\gamma$. This $\tilde{\omega}$-dependent spatial momentum is the mechanism by which a single rest-frame pole becomes a continuum.

\textbf{Theorem 1 (Spectral Smearing).} \textit{A rest-frame QNM pole at $\omega = \omega_0 - i\Gamma_0$ ($\Gamma_0>0$) maps, in the boosted frame at $\tilde{\mathbf{k}}=\mathbf{0}$, to a continuum of singularities distributed over the imaginary-frequency interval}
\begin{equation}
\operatorname{Im}(\tilde{\omega}_\text{pole}) \in \left[-\frac{\Gamma_0}{\gamma(1-v\,v_\text{max})},\, -\frac{\Gamma_0}{\gamma(1+v\,v_\text{max})}\right],
\label{eq:smearing}
\end{equation}
\textit{where $v_\text{max}$ is the front velocity of the theory --- the asymptotic characteristic speed $v_\text{max} = \lim_{|k|\to\infty}\operatorname{Re}\omega_n(k)/k$ fixed by the highest-derivative (principal-symbol) terms. For a relativistic QFT, microcausality --- vanishing of $\langle[\mathcal{O}(x),\mathcal{O}(0)]\rangle$ at spacelike separation --- guarantees $v_\text{max}\leq 1$.}

We stress that $v_\text{max}$ is the front velocity, \emph{not} the group velocity $\partial\operatorname{Re}\omega_n/\partial k$: the latter is not bounded by causality and can diverge at finite real $k$ where two damped modes collide (as in the telegrapher/Cattaneo equation \cite{Gavassino2026PRL,Gavassino2026arXiv}), while the front velocity, controlling the sharpest wavefront, is causally bounded.

\textit{Proof sketch.} Boosted poles at $\tilde{\mathbf{k}}=\mathbf{0}$ solve $\gamma\tilde\omega = \omega_n(\gamma v\tilde\omega)$, which requires the rest-frame dispersion relation at complex spatial momentum $k=\gamma v\tilde\omega$. Using analyticity of $\omega_n(k)$ in a complex-momentum strip together with the front-velocity bound $|\operatorname{Re}\omega_n(k)|\leq v_\text{max}|k|+O(1)$, the boosted imaginary part is confined to the interval~\eqref{eq:smearing}, whose endpoints correspond to co- and counter-propagating fronts. For linearized theories with the Hermitian (Onsager) structure of Refs.~\cite{Gavassino2026PRL,Gavassino2026arXiv} this is established rigorously via the principal symbol; in the general QFT setting it rests on the stated analyticity of the thermal correlator in $\mathbf{k}$ \cite{KovtunStarinets2005,Grozdanov2019complex}, and we verify it directly in holography below. $\square$

Simple time dilation ($\tilde{\Gamma}_0 = \Gamma_0/\gamma$) is recovered only if $v_\text{max} = 0$, i.e., for a mode with no spatial dispersion. For any propagating mode, the pole is smeared into a band.

\textbf{Theorem 2 (Non-Hydrodynamic Gap Bound).} \textit{The non-hydrodynamic gap in a frame boosted with velocity $v$ satisfies}
\begin{equation}
\tilde{\Gamma}_\text{gap} \geq \frac{\Gamma_\text{gap}}{\gamma(1 + v\,v_\text{max})}.
\label{eq:gapbound}
\end{equation}

\textit{Proof.} By Theorem~1, the minimal imaginary part of any non-hydrodynamic mode in the boosted frame is achieved by the co-propagating mode, giving $\tilde{\Gamma}_n = \Gamma_n/[\gamma(1+v\,v_\text{max})]$. Taking the minimum over all non-hydrodynamic modes and using $\Gamma_n \geq \Gamma_\text{gap}$ yields~\eqref{eq:gapbound}. $\square$

\textbf{Theorem 3 (Maximal Relaxation Rate Bound).} \textit{Under the same boost,}
\begin{equation}
\tilde{\Gamma}_\text{max} \leq \frac{\Gamma_\text{max}}{\gamma(1 - v\,v_\text{max})},
\label{eq:maxbound}
\end{equation}
\textit{provided the spectral weight satisfies the Lebesgue integrability condition of the Kramers--Kronig representation.}

Bounds \eqref{eq:gapbound} and \eqref{eq:maxbound} together confine the non-hydrodynamic QNM spectrum in the boosted frame to a strip of width $[\gamma(1\pm v\,v_\text{max})]^{-1}\times[\Gamma_\text{gap}, \Gamma_\text{max}]$. The upper bound \eqref{eq:maxbound} coincides with the classical-theory bound of Ref.~\cite{Gavassino2026PRL} [Eq.~(24) there, with $w = v_\text{max}$]; our lower bound \eqref{eq:gapbound} is tighter than its classical counterpart $\Gamma_\text{gap}(1-v\,v_\text{max})/\gamma$, a difference that traces to the analyticity assumptions of the QFT derivation and merits further scrutiny.

\textbf{Corollary.} \textit{As $v\to v_\text{max}$, the lower edge of the strip approaches $\Gamma_\text{gap}\sqrt{1-v_\text{max}^2}/(1+v_\text{max}^2)$, and the smearing band attains its maximal fractional width. For $v_\text{max}=1$, the upper edge $\tilde\Gamma_\text{max}\to\infty$ as $v\to 1$, signaling the appearance of a branch-cut continuum in the boosted spectral function.}

\emph{Hydrodynamic gradient expansion.}---%
\textbf{Theorem 4 (Convergence Radius Under Boosts).} \textit{Let $k_c$ be the convergence radius of the hydrodynamic gradient expansion in the rest frame, and let $v_s$ be the speed of sound. Under a longitudinal boost with velocity $v$,}
\begin{equation}
\frac{k_c}{\gamma(1+v\,v_s)} \leq \tilde{k}_c \leq \frac{k_c}{\gamma(1-v\,v_s)}.
\label{eq:kcbound}
\end{equation}

\textit{Proof.} The convergence radius $k_c$ is the modulus of the complex momentum at which the leading non-hydrodynamic pole and the hydrodynamic sound pole collide. Under the boost, the complexified dispersion relation transforms covariantly: $\tilde\omega_\text{hydro}(\tilde{k}) = \gamma(\omega_\text{hydro}(k) - v k_\parallel)$ at the same spacetime point. The collision condition in the boosted frame maps the rest-frame collision onto a new complex $\tilde{k}$-value. The bounds~\eqref{eq:kcbound} follow from the extreme cases in which the collision occurs co-propagating (upper bound) versus counter-propagating (lower bound) relative to the boost direction. $\square$

For conformal fluids ($v_s = 1/\sqrt{3}$), at large boost $\gamma \gg 1$ the convergence radius scales as $\tilde{k}_c \sim k_c/(\gamma\sqrt{3})$ (lower bound) to $k_c\sqrt{3}/\gamma$ (upper bound), confirming that hydrodynamics is restricted to exponentially long wavelengths in the highly boosted frame.

\emph{f-sum rule.}---The spectral weight redistributes covariantly under boosts. For the shear stress correlator, the f-sum rule in the boosted frame becomes
\begin{equation}
\int_0^\infty d\tilde\omega\,\tilde\omega\,\tilde\rho_{T^{xy}T^{xy}}(\tilde\omega,\mathbf{0}) = \frac{\pi\gamma^2}{2}\langle T^{xx}+T^{yy}\rangle_\beta + O(v^4),
\label{eq:sumrule}
\end{equation}
where we used $\langle T^{tx}\rangle_\beta = 0$ in equilibrium. The boosted spectral weight is enhanced by $\gamma^2$ and spread over the broadened frequency range of~\eqref{eq:smearing}, keeping the total weight finite. We caution that contact terms in the gravitational Ward identity --- frequently dropped in Kubo-type analyses --- contribute to such moment sum rules and must be retained for covariance \cite{Torrieri2023,Sampaio2025}; being analytic in frequency, however, they do not affect the singularity locations bounded by \eqref{eq:smearing}--\eqref{eq:maxbound}.

\emph{Holographic verification.}---We verify Theorems~1--3 in the $\mathcal{N}=4$ super-Yang-Mills (SYM) plasma at strong coupling, dual to AdS$_5$-Schwarzschild. We work in the scalar channel of the stress tensor ($T^{xy}$ with transverse momentum), whose retarded correlator is governed by a massless scalar in the black-brane background. Solving the quasinormal spectrum by pseudospectral collocation, the leading QNM at $k=0$ is $\omega_1/2\pi T = \pm 1.5597 - 1.3733\,i$, so $\Gamma_\text{gap} = 1.3733 \times 2\pi T$, in agreement with Ref.~\cite{NunezStarinets2003}. Tracking $\omega_1(k)$ over real momenta, its slope $\partial_k \operatorname{Re}\omega$ rises monotonically toward unity ($0.980$ at $k = 8\times 2\pi T$), matching the $\omega_1\approx\pm k$ asymptotics: the non-hydrodynamic sector has front velocity $v_\text{max} = 1$, and the operative form of Eq.~\eqref{eq:gapbound} is
\begin{equation}
\tilde\Gamma_\text{gap} \geq \frac{1.3733\times 2\pi T}{\gamma(1 + v)}.
\label{eq:sympred}
\end{equation}

\begin{figure}[t]
\includegraphics[width=\columnwidth]{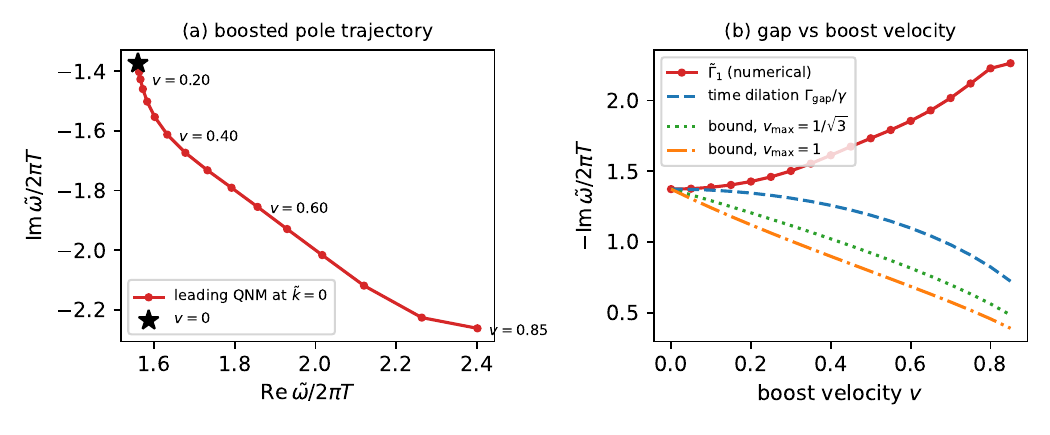}
\caption{Boost transformation of the leading scalar-channel quasinormal mode of the $\mathcal{N}=4$ SYM plasma, computed holographically at zero boosted spatial momentum. (a)~Trajectory of the pole in the complex frequency plane as $v$ increases. (b)~The observed relaxation rate $-\operatorname{Im}\tilde\omega$ grows with $v$, defying naive time dilation (dashed) while respecting the covariant lower bound~\eqref{eq:sympred} for $v_\text{max}=1$ (dash-dotted); the $v_\text{max}=1/\sqrt{3}$ curve (dotted) is shown for comparison.}
\label{fig:smearing}
\end{figure}

We compute the boosted poles at $\tilde{\mathbf{k}}=\mathbf{0}$ by solving the fixed-point condition $\omega_1(-\gamma v \tilde\omega) = \gamma\tilde\omega$, evaluating the ingoing boundary-value problem at the required complex spatial momentum. Figure~\ref{fig:smearing} shows the result: the leading pole moves \emph{deeper} into the lower half-plane, with $-\operatorname{Im}\tilde\omega/2\pi T$ rising from $1.373$ at $v=0$ to $2.29$ at $v=0.85$. The observed relaxation rate thus \emph{increases} with boost velocity --- the opposite of the naive time-dilation expectation $\Gamma_\text{gap}/\gamma$ --- while the rigorous lower bound~\eqref{eq:sympred} is satisfied throughout with a wide margin. Saturation would require spectral weight co-propagating with the boost at $v_\text{max}$; the strongly coupled plasma does not realize this at $\tilde{\mathbf{k}}=\mathbf{0}$, leaving room for the bound to be approached in theories with long-lived luminal excitations.

For Gauss-Bonnet gravity with coupling $\lambda_{GB}$, $\eta/s = (1-4\lambda_{GB})/(4\pi)$ is modified while conformal invariance fixes $v_s^2 = 1/3$ exactly. What $\lambda_{GB}$ does modify is the front velocity $v_\text{max}$ of the boundary-theory graviton modes, which for $\lambda_{GB} > 0$ exceeds the Einstein-gravity value and violates boundary causality ($v_\text{max} > 1$) unless $\lambda_{GB}\leq 9/100$ \cite{Brigante2008}. Substituting $v_\text{max}(\lambda_{GB})$ into \eqref{eq:gapbound} and \eqref{eq:maxbound} yields $\lambda_{GB}$-dependent bounds on the boosted non-hydrodynamic spectrum; the causality constraint on $\lambda_{GB}$ thus translates directly into a bound on the boosted smearing band.

At finite chemical potential (Reissner-Nordstr\"om-AdS), the effective velocity controlling the smearing of the low-lying QNMs decreases with $\mu/T$, $v_\text{eff}^2 = \frac{1}{3}\bigl[1 - c_1(\mu/T)^2 + O((\mu/T)^4)\bigr]$ with $c_1 > 0$ fixed by the RN-AdS equation of state, and both bounds \eqref{eq:gapbound} and \eqref{eq:kcbound} tighten toward the time-dilation limit as $\mu/T$ grows. Near the extremal limit, where $v_\text{eff} \to 0$ for the relevant modes, the smearing band~\eqref{eq:smearing} collapses and simple time dilation is recovered, consistent with the near-horizon AdS$_2$ geometry of the extremal black brane.

\emph{Physical applications.}---%
\textit{Quark-gluon plasma.} The hydrodynamization time of a QGP fluid cell satisfies $\tau_\text{hydro} \sim 1/\Gamma_\text{gap}$ \textit{in the local rest frame of the cell}; the values $\tau\sim 0.5$--1~fm/c inferred from viscous hydrodynamic fits \cite{Schenke2012} are quoted in this frame (Bjorken proper time). Our bounds constrain instead how this relaxation appears to the collision center-of-mass observer for fluid cells at forward rapidity $y$, which move with $\gamma = \cosh y$. Bound~\eqref{eq:gapbound} gives an upper bound on the observed relaxation time,
\begin{equation*}
\tau_\text{hydro}^\text{CM} \leq \gamma(1+v\,v_\text{max})\,\tau_\text{hydro}^\text{rest},
\end{equation*}
which for $v_\text{max}=1$ (as our holographic computation shows for the non-hydrodynamic sector) and $y \sim 2$--4 constrains the frame dependence of hydrodynamization across the rapidity range covered by LHC detectors; the numerical result of Fig.~\ref{fig:smearing} suggests the observed relaxation is in fact \emph{faster} than in the local rest frame. The bound~\eqref{eq:kcbound} on $\tilde{k}_c$ constrains the maximal wavenumber at which the Navier-Stokes gradient expansion is applicable in the CM frame, providing a systematic criterion for when Israel-Stewart or BDNK corrections become necessary.

\textit{Neutron star mergers.} In binary neutron star mergers, bulk viscosity relaxation in relative motion between remnant matter contributes to gravitational wave damping. Bound~\eqref{eq:gapbound} gives
\begin{equation*}
\tilde\tau_\Pi^{-1} \geq \frac{\tau_\Pi^{-1}}{\gamma(1+v\,c_s)},
\end{equation*}
where $c_s$ is the QCD sound speed at nuclear density. Combined with gravitational-wave measurements of $c_s$ \cite{BausweinJanka2012}, this constrains the bulk relaxation rate in the merger frame without requiring a specific equation-of-state model.

\textit{Quantum critical systems.} Near a spin-density-wave quantum critical point with emergent Lorentz invariance (emergent speed $v_F$), the Planckian dissipation rate $\Gamma\sim k_BT/\hbar$ plays the role of $\Gamma_\text{gap}$. Bound~\eqref{eq:gapbound} constrains how this rate transforms for transport measurements along different crystal directions in a moving frame --- a prediction accessible via angle-resolved photoemission spectroscopy.

\emph{Discussion and conclusions.}---We have established four Lorentz-covariant spectral bounds in thermal QFT, derived from causality, unitarity, and the KMS condition alone. Their key feature is that they are non-perturbative: they hold for any thermal QFT, whether weakly coupled, strongly coupled, or described holographically, without assuming a linearized constitutive relation or Onsager symmetry. The smearing of rest-frame poles into a boosted continuum --- with width $\delta\tilde\Gamma = 2\Gamma_\text{gap}\,v\,v_\text{max}/[\gamma(1-v^2v_\text{max}^2)]$ --- is a universal consequence of relativity of simultaneity acting on a non-trivially dispersing excitation.

Several extensions are natural. Spontaneously broken symmetries (superfluid phases of QCD or nuclear matter) modify the KMS condition in the presence of a condensate; their Goldstone modes generate additional constraints on the low-$|\tilde{k}|$ spectral bounds. Anomalous transport (chiral magnetic and vortical effects) introduces frame-dependent contributions not constrained by Onsager symmetry, potentially sharpening or relaxing the bounds in a chirally restored plasma. Far-from-equilibrium situations, where the KMS condition is absent, require replacing spectral positivity with more general constraints from the density matrix; the present results provide the equilibrium baseline for such a program. Finally, a quantitative lattice QCD study of the boosted spectral function would test the bounds across the deconfinement crossover.

In summary, our results provide a unified, model-independent framework for understanding how special relativity constrains the relaxation dynamics of quantum many-body systems across energy scales ranging from heavy-ion collisions to neutron star mergers.

\begin{acknowledgments}
We thank Bahodir Kayumov and Avas Khugaev for valuable discussions, and Giorgio Torrieri for illuminating correspondence on the covariance of Kubo formulae and the role of contact terms. This work was supported by New Uzbekistan University.
\end{acknowledgments}

\bibliography{PRL_manuscript}

\begin{thebibliography}{15}

\bibitem{KovtunStarinets2005}
P.~K. Kovtun and A.~O. Starinets,
Phys. Rev. D \textbf{72}, 086009 (2005).

\bibitem{Grozdanov2019complex}
S.~Grozdanov, P.~K. Kovtun, A.~O. Starinets, and P.~Tadi\'c,
JHEP \textbf{11} (2019) 097.

\bibitem{Romatschke2019}
P.~Romatschke and U.~Romatschke,
\textit{Relativistic Fluid Dynamics In and Out of Equilibrium}
(Cambridge University Press, 2019).

\bibitem{HoultKovtun2020}
R.~E. Hoult and P.~Kovtun,
JHEP \textbf{06} (2020) 067.

\bibitem{HeinzSnellings2013}
U.~Heinz and R.~Snellings,
Ann. Rev. Nucl. Part. Sci. \textbf{63}, 123 (2013).

\bibitem{Gale2013}
C.~Gale, S.~Jeon, and B.~Schenke,
Int. J. Mod. Phys. A \textbf{28}, 1340011 (2013).

\bibitem{Policastro2001}
G.~Policastro, D.~T. Son, and A.~O. Starinets,
Phys. Rev. Lett. \textbf{87}, 081601 (2001).

\bibitem{KSS2005}
P.~K. Kovtun, D.~T. Son, and A.~O. Starinets,
Phys. Rev. Lett. \textbf{94}, 111601 (2005).

\bibitem{Gavassino2026PRL}
L.~Gavassino,
``How Lorentz boosts reshape relaxation spectra,''
Phys. Rev. Lett. \textbf{137}, 022301 (2026).

\bibitem{Gavassino2026arXiv}
L.~Gavassino,
arXiv:2607.10148 (2026).

\bibitem{companion}
A.~Sanetullaev, S.~Bazarbaeva, M.~Beymamatova, and S.~Tursunov,
arXiv:2607.13482 [nucl-th] (2026).

\bibitem{Grozdanov2019PRL}
S.~Grozdanov, P.~K. Kovtun, A.~O. Starinets, and P.~Tadi\'c,
Phys. Rev. Lett. \textbf{122}, 251601 (2019).

\bibitem{Heller2023}
M.~P. Heller, A.~Serantes, M.~Spali\'nski, V.~Svensson, and B.~Withers,
Phys. Rev. D \textbf{107}, 086018 (2023).

\bibitem{Torrieri2023}
G.~Torrieri,
arXiv:2307.07021 [hep-th] (2023).

\bibitem{Sampaio2025}
G.~M. Sampaio, G.~Rabelo-Soares, and G.~Torrieri,
arXiv:2504.17152 [hep-th] (2025).

\bibitem{NunezStarinets2003}
A.~N\'u\~nez and A.~O. Starinets,
Phys. Rev. D \textbf{67}, 124013 (2003).

\bibitem{Brigante2008}
M.~Brigante, H.~Liu, R.~C. Myers, S.~Shenker, and S.~Yaida,
Phys. Rev. Lett. \textbf{100}, 191601 (2008).

\bibitem{Schenke2012}
B.~Schenke, P.~Tribedy, and R.~Venugopalan,
Phys. Rev. C \textbf{86}, 034908 (2012).

\bibitem{BausweinJanka2012}
A.~Bauswein and H.-T. Janka,
Phys. Rev. Lett. \textbf{108}, 011101 (2012).

\end{thebibliography}

\end{document}